\documentclass[twocolumn,amsmath,showpacs,preprintnumbers,pra]{revtex4}
\usepackage[dvips]{graphicx}
\newcommand{\tcoin}{\ensuremath{\tau_c}}
\newcommand{\Ratessi}{\ensuremath{P_{ssi}}}
\newcommand{\gtwo}{\ensuremath{g^{(2)}}}
\newcommand{\bargtwo}{\ensuremath{{\bar g}^{(2)}}}
\newcommand{\gsi}{\ensuremath{\gtwo_{si}}}
\newcommand{\gc}{\ensuremath{\gtwo_{c}}}
\newcommand{\Nsi}{\ensuremath{N_{si}}}
\newcommand{\Nthree}{\ensuremath{N_{ssi}}}
\newcommand{\gsiav}{\ensuremath{\bargtwo_{si}}}
\newcommand{\gcav}{\ensuremath{\bargtwo_{c}}}
\begin{document}
\title{Comment on ``Coherence measures for heralded single-photon sources''}
\author{Stefano Bettelli}
\affiliation{Romanogasse 29/15, 1200 Wien, \"Osterreich}
\preprint{arXiv:0911.1478}

\begin{abstract}
A recent Brief Report [Phys. Rev. A {\bf 79}, 035801 (2009)] investigates two
figures of merit for her\-alded photon sources based on spontaneous parametric
down-conversion with a continuous pump, namely the time-averaged temporal
coherence between the signal and idler beams and the time-averaged {\em
  conditioned} second-order degree of coherence for photons in the heralded
arm. However, contrary to what is claimed, no one-arm second-order effect is
actually measured.
\end{abstract}

     %% PDC and production of entangled photons

\maketitle

The phenomenon of spontaneous parametric down-conversion (SPDC) consists in the
splitting of a pump photon, in a non-linear crystal, into two modes (arms). A
detection in one arm (idler) implies the presence of a photon in the other
(signal): the signal photon can thus be heralded in a time window of width
$\tau_d$, set by the indetermination on the detection time. The probability
that a second signal photon shows up in the window is, for sufficiently large
windows, essentially $R\tau_d$, where $R$ is the pair-generation rate, which
can be made as low as desired. For this reason SPDC-based sources are often
used to emulate single-photon sources. 

The apparent Poissonian statistics in one arm of the SPDC source\footnote{The
  statistical properties described in this paragraph are discussed at length in
  the introduction of Ref. \cite{Blauensteiner09}.} observed in the heralded
time window is determined by the fact that $\tau_d$ is often much larger than
$\Delta t$, the signal coherence time, typically in the range $0.1$ps to $1$ps;
only when $\tau_d \sim \Delta t$ the thermal nature of SPDC radiation shows up,
and the probability of a second photon in the window approaches $2R\tau_d$.
Experiments on SPDC intrinsic one-arm statistics in single-photon regime are
difficult to perform when sources are operated with a continuous pump, since in
this case the jitter $\tau_d$ of single-photon detectors is not better than
$\sim 10^2$ps. Thermal statistics for one arm of a SPDC source operated in this
way was demonstrated only recently \cite{Blauensteiner09}.

A recent Brief Report by Bocquillon {\em et al.}~\cite{Bocquillon09} (see also
the expanded version by Razavi {\em et al.}~\cite{Razavi09}), concerning the
characterisation of heralded SPDC-based photon sources with a continuous pump,
claims ``{\em \dots [to carve] into the theoretical aspects of [measuring] the
  temporal correlation between the signal and idler beams, and \dots~the
  second-order degree of coherence for the heralded signal photons}~\rlap{.}''
More specifically, since the time indetermination of the detection apparatus is
orders of magnitude larger than the coherence times of the source, the authors
develop a theory connecting the values of the parameters intrinsic to the
investigated system to the actually observed averages, and test the theory
experimentally. In particular, in the second part of the report, the authors
investigate the detection of two photons in the signal arm conditioned on the
detection of one photon in the idler arm; clearly this measurement is strictly
related to that described in \cite{Blauensteiner09}, and definitely not less
difficult. However, contrary to what is claimed, no one-arm second-order effect
is actually measured in \cite{Bocquillon09}, as shown in the following.

\section*{Preliminaries}

\begin{figure}[b!]
\includegraphics[width=.99\linewidth]{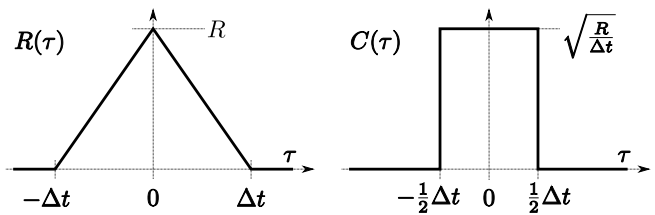}
\caption{\label{fig:plotRC}The signal-signal or idler-idler
  autocorrelation function $R(\tau)$, and the signal-idler
  cross-correlation function $C(\tau)$ for the SPDC process analysed
  in \cite{Bocquillon09}. $R$ is the pair-generation rate, while $1 /
  \Delta t$ is the bandwidth. Vertical scales are very different since
  in the actual experiment $R\Delta t \sim 1.4 \cdot 10^{-5}$.}
\end{figure}

\begin{figure}[t!]
\includegraphics[width=.99\linewidth]{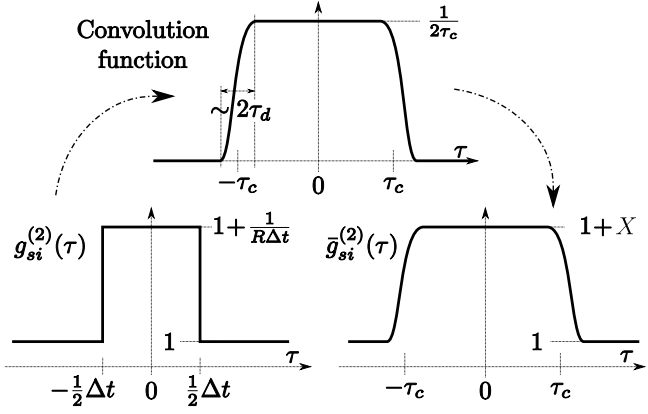}
\caption{\label{fig:g2si} \label{fig:g2si_av} The second-order coherence
  function between signal and idler, $\gsi(\tau)$, and its time average,
  $\gsiav(\tau)$. The convolution function that transforms the former into the
  latter is represented on top. Since $X = (2R\tcoin)^{-1} \sim 10^1$, the
  smeared signal-idler correlation peak, though strongly reduced with respect
  to the ideal one, is still rather simple to measure.}
\end{figure}

The authors of \cite{Bocquillon09} assume a theoretical model for a low-gain
frequency-degenerate type-II SPDC process predicting that all auto- and
cross-correlations of the source arms can be expressed by means of the two real
even functions $R(\tau)$ and $C(\tau)$ shown in Fig.\ref{fig:plotRC}. In
practice $R(\tau)$ is the first-order coherence of both arms, and $R(0) = R
\sim 43$MHz ($R_{\mathrm{SPDC}}$ in the original text) is the pair-generation
rate. With the help of the Gaussian moment-factoring theorem the temporal
correlation \smash{$\gsi(\tau)$} of the signal and the idler,
Fig.\ref{fig:g2si}, is shown to be sharply peaked for $|\tau| < \Delta t/2$
(the parameter $\Delta t \sim 0.3$ps plays the role of a signal-idler coherence
time):
\begin{align}
  \gsi(\tau) \label{eq:value_gsi}
  &= \frac{\langle E_s^\dagger(t + \tau) 
    E^\dagger_i(t) E_i(t) E_s(t + \tau) \rangle}
  {\langle E^\dagger_s(t + \tau) E_s(t + \tau) \rangle \cdot
    \langle E^\dagger_i(t) E_i(t) \rangle} \notag \\
  &= 1 + \frac{C^2(\tau)}{R^2(0)} =
  \begin{cases}
    1 + \frac{1}{R\Delta t} & \textrm{if~}|\tau| < \frac{\Delta t}{2} \\
    1 & \textrm{otherwise}
  \end{cases}
\end{align}

This sharp peak is smeared by detector response times and by the finite
coincidence window (implemented in software), respectively characterised by
$\tau_d$ and $\tcoin$ ($\tau_{\mathrm{coin}}$ in the original text). The
authors of \cite{Bocquillon09} model these effects by means of convolution
functions. For the purposes of this Comment it is sufficient to consider
$\tcoin \gg \tau_d$; in this case the combined response function of the
hardware-software setup is equivalent to a moving-window average with size
$2\tcoin$ and transition regions as large as $2\tau_d$. The time-averaged
second-order coherence $\gsiav(\tau)$, proportional to the measured rate of
signal-idler coincidences $\Nsi(\tau)$, will be equal to one for $|\tau| >
\tcoin + \tau_d$, and to a constant when $|\tau| < \tcoin-\tau_d$, {\em i.e.},
when the large peak of \smash{$\gsi(\tau)$} falls in the central part of the
convolution window,
\begin{equation} \label{eq:g2si}
  \gsiav(|\tau| < \tcoin - \tau_d) 
%  = \frac{1}{2\tcoin} \int_{-\tcoin}^{+\tcoin} d\tau\, \gsi(\tau)
  = 1 + \frac{1}{2R\,\tcoin} \stackrel{\mathrm{def}}{=} 1 + X,
\end{equation}
where $X=(2R\,\tcoin)^{-1}$ is a signal-idler coincidence factor. Since $R \sim
43$MHz and $2\tcoin < 3$ns, the additional term in Eq.\ref{eq:g2si} is still of
the order of $10^1$, and therefore the signal-idler temporal correlation is
relatively easily to demonstrate. Of course, this signal-idler peak could be
made arbitrary large by reducing the rate $R$ (in practice, by reducing the
intensity of the pump, as in \cite[Fig.4]{Razavi09}). At the same time, given
that both time scales of the experimental convolution function, $\tau_d \sim
350$ps and $\tcoin \gtrsim 0.5$ns, are several orders of magnitude larger than
the coherence time $\Delta t \sim 0.3$ps, the experiment is only sensitive to
the integral of the peak, and not to a specific peak model. The behaviour of
the average $\gsiav(\tau)$ is illustrated in Fig.\ref{fig:g2si_av}, and agrees
with \cite[Fig.~2]{Bocquillon09} (also quantitatively, but for the curve at
$2\tcoin = 0.78$ns for which $\tau_d \not\!\ll \tcoin$).

The authors of the Brief Report claim that good agreement between theory and
experimental data proves that their model is particularly accurate, but indeed
it is only the dependence on $\tcoin$ that is truly confirmed, and repeating
data analysis with different values is just a test that the software is working
correctly.

\section*{Measurement of the conditioned second-order coherence}

In the second part of the Brief Report the authors \mbox{analyse} the
statistics of signal-signal coincidences conditioned on idler-photon detection,
defined as
\begin{equation} \label{eq:def_gc}
  \gc(t_1,t_2|t_i) = 
  \frac{\langle E_s^\dagger(t_1) E_s^\dagger(t_2) E_s(t_2) E_s(t_1) \rangle_{pm}}
       {\langle E_s^\dagger(t_1) E_s(t_1) \rangle_{pm}
       \langle E_s^\dagger(t_2) E_s(t_2) \rangle_{pm}},
\end{equation}
where the $\langle \dots \rangle_{pm}$ averages are performed on
post-measurement states ({\em i.e.}, after detection of an idler photon). By
using the operator identity $\langle Y \rangle_{pm} = \langle E_i^\dagger(t_i)
Y E_i(t_i) \rangle \cdot \langle E_i^\dagger(t_i) E_i(t_i) \rangle^{-1}$,
Eq.\ref{eq:def_gc} is reduced to
\begin{equation} \label{eq:formula_gc}
  \gc(t_1,t_2|t_i) = \frac{\Ratessi(t_1,t_2,t_i)}
     {R^3 \cdot \gsi(t_1-t_i) \cdot \gsi(t_2-t_i)},
\end{equation}
where $\Ratessi$ is the ideal triple-coincidence rate that, again with the help
of the Gaussian moment-factoring theorem, is proven to be equal to
{\renewcommand{\min}{\!-\!}
  \begin{multline} \label{eq:formula_Ratessi}
    \hspace{-2ex} \Ratessi(t_1,t_2,t_i)
    = 2 C(t_1\min t_i) C(t_2\min t_i) R(t_1\min t_2) + \\ + R \left[R^2 +
      R^2(t_1\min t_2) + C^2(t_1\min t_i) + C^2(t_2\min t_i) \right].
\end{multline}}
\begin{figure}[b!]
\includegraphics[width=.99\linewidth]{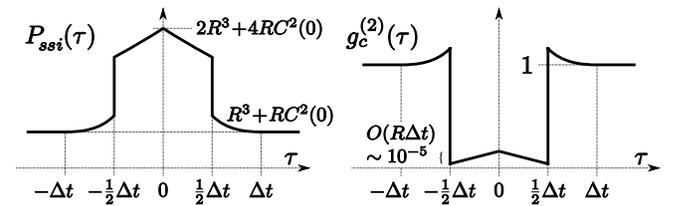}
\caption{\label{fig:g2c} The triple-coincidence rate $\Ratessi(\tau)$, and the
  second-order correlation function $\gc(\tau)$ for an idler photon arriving at
  time $t_i$ and two signal photons arriving at $t_i$ and $t_i+\tau$, as
  defined by \cite[Eq.9]{Bocquillon09}. The temporal correlation function is
  very small ($\sim 10^{-5}$) in the region $\tau \in [-\Delta t/2, \Delta
    t/2]$.}
\end{figure}
Actual measurements are performed with the particular case $t_1=t_i$ and
$t_2=t_i+\tau$ in mind, so that the relevant formula for the numerator of
Eq.\ref{eq:formula_gc} is
\begin{multline}
  \Ratessi(\tau) \stackrel{\scriptscriptstyle \mathrm{def}}{=}
  \Ratessi(t_i,t_i+\tau, t_i) = \\ \quad = 2 C(0) C(\tau) R(\tau)
  + R \left[ R^2 \!+\! R^2(\tau) \!+\! C^2(0) 
    \!+\! C^2(\tau) \right]. \raisetag{7ex}
\end{multline}
This expression cannot be simplified further, but its behaviour is easy to
understand. For delays larger than the coherence time all non-constant terms
die out, and $\Ratessi \rightarrow R^3 + R C^2(0) = R^3 \gsi(0)$; for $|\tau| <
\Delta t$ the shape is peaked, and its maximum value does not exceed four times
its large-delay value, as shown in Fig.\ref{fig:g2c}. The denominator of
Eq.\ref{eq:formula_gc}, {\em i.e.} $R^3 \gsi(0) \,\gsi(\tau)$, is almost always
\smash{$R^3 \gsi(0)$}, but in the narrow region $[-\Delta t/2, \Delta t/2]$ it
grows several orders of magnitude to reach $R^3 [ \gsi(0) ]^2$, as can be seen
in Fig.\ref{fig:g2si}. Therefore the function of interest, $\gc(\tau) =
\gc(t_i,t_i+\tau|t_i)$, represented in Fig.\ref{fig:g2c}, has unit value for
$|\tau|>\Delta t$, and is $O(R\Delta t) \sim 10^{-5}$ for $|\tau|<\Delta t/2$.

\begin{figure}[t!]
\includegraphics[width=.99\linewidth]{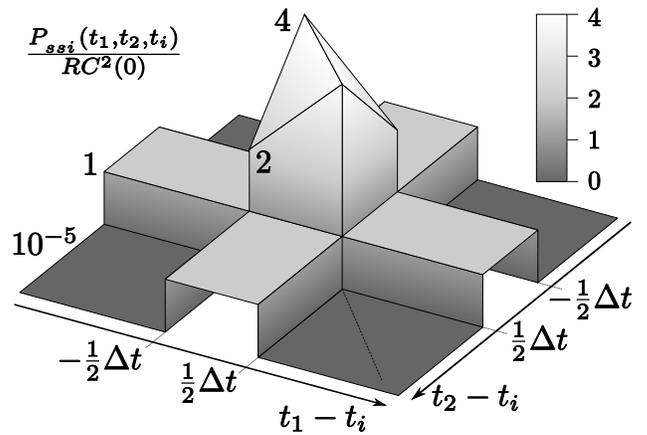}
\caption{\label{fig:Pssi} The ideal triple-coincidence rate
  $\Ratessi(t_1,t_2,t_i)$ (only most relevant structures shown). The ridges
  running parallel to the axes $t_1=t_i$ and $t_2=t_i$ are given by
  signal-idler correlation. When both $t_1$ and $t_2$ are very close to $t_i$
  correlations are increased by a factor at most four.  $\Ratessi$ is symmetric
  under $t_1 \leftrightarrow t_2$, i.e., around the dash-dotted bisector. The
  function floor is the accidental triple rate $R^3$.}
\end{figure}

Experimentally measuring a time-averaged version of this well is far from
trivial, since the size of the well is only $O(\Delta t)$, and the typical
convolution spread due to detector jitter and finite coincidence windows is
$\tcoin$, so that the effect is in the best case $O(\Delta t / \tau_d) \sim
10^{-3}$, and even smaller with the coincidence windows used in the
article. This is an effect of the same order of magnitude as that measured in
\cite{Blauensteiner09} on an almost identical physical system. Instead, Fig.3
of \cite{Bocquillon09} shows a large well extending over $1$ns or more. The
reason for this strange result is that the authors actually plot the
``time-averaged second-order correlation function'' $\gcav(\tau)$ defined as
\begin{equation} \label{eq:defgcav}
  \gcav(\tau) = \frac{\Nthree(\tau) \cdot R}{\Nsi(0) \cdot \Nsi(\tau)},
\end{equation}
where $\Nsi$ is the already mentioned signal-idler coincidence rate, and
$\Nthree$ is the triple signal-signal-idler coincidence rate, proportional to a
smeared \smash{$\Ratessi$}. It is important to understand that even if $\gcav$
is studied for $t_1=t_i$, the observable $\Nthree$ is the outcome of a
triple-coincidence experiment, and its value is linked to the full three-time
structure of $\Ratessi$, shown in Fig.\ref{fig:Pssi}. Minor details apart,
$\Ratessi(t_1,t_2,t_i)$ is characterised by three levels: the accidental
triple-coincidence plateau at $R^3$, the signal-idler coincidence ridges at
$R^3 + RC^2(0)$ (when $|t_1-t_i| < \Delta t/2$ or $|t_2-t_i| < \Delta t/2$),
and a more complicated central peak (when both $t_1, t_2 \in [t_i - \Delta
  t/2,t_i + \Delta t/2]$) where the correlation grows even more, up to a factor
four.

The experimental rate $\Nthree$ is obtained by convoluting $\Ratessi$ with the
response functions of three detectors and two coincidence windows. The effect
can be modelled as a moving average with a square bidimensional averaging
window of area $(2\tcoin)^2$ and transition regions of size $2t_d$. The value
of $\Nthree$ will then be significantly different from $R^3$ only when the
averaging window is close to one of the ridges of Fig.\ref{fig:Pssi}; due to
translation invariance the result is $R^3 [1 + X]$ as before. But when the
averaging window sits at $t_1 = t_2 = t_i$ both ridges participate to the
average and their contribution is twice as large: $R^3 [1 + 2X]$. Note that the
complicated peak at the ridge junction is almost irrelevant since its
contribution relative to $R^3 X$ is only $O(\Delta t/\tcoin) \sim 10^{-3}$, due
to its height being limited and its base being only $\Delta t \times \Delta t$,
and not $\Delta t \times \tcoin$ as for the part of each ridge within the
averaging window.

\begin{figure}[t!]
\includegraphics[width=.99\linewidth]{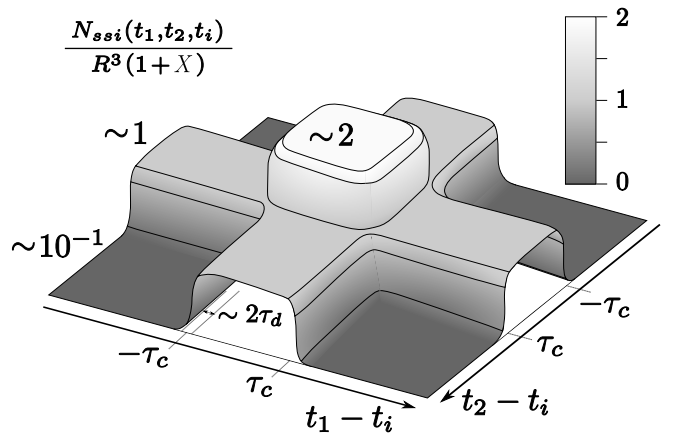}
\caption{\label{fig:Nssi} The expected experimental triple-coincidence rate
  $\Nthree(t_1,t_2,t_i)$, obtained from the ideal triple-coincidence rate
  $\Ratessi(t_1,t_2,t_i)$ of Fig.\ref{fig:Pssi} by means of a moving
  bidimensional averaging window of area $4\tcoin^2$ and transition regions of
  size $2\tau_d$. The central structure is in practice not related to the
  central peak of Fig.\ref{fig:Pssi}.  The scales of time axes in
  Fig.\ref{fig:Pssi} and \ref{fig:Nssi} differ by more than three orders of
  magnitude. The floor of the function is again the accidental triple rate
  $R^3$.}
\end{figure}

The conclusion of the previous discussion is that the expected shape of the
coincidence functions \footnote{A plot of $\Nthree$ is not available in
  Ref.\cite{Bocquillon09}, but it is given in \cite[Fig.~5]{Razavi09}, and it
  corresponds to the analysis just given (see Fig.\ref{fig:Nssi} here). The
  authors of \cite{Razavi09} claim that the peak at the centre of
  \cite[Fig.~5]{Razavi09} represents the contribution of multiple photon pairs,
  but this is clearly wrong, since it is just an artifact of the averaging
  method. In fact, the peak height is twice as large as the height along the
  ridges, and not more as should be expected from Fig.\ref{fig:Pssi} (the
  authors blame this discrepancy on the 50/50 beam-splitter, but this
  explanation is incorrect since the beam-splitter influences all rates in the
  same way).} are
\begin{align}
  \Nthree(\tau) &\sim
  \begin{cases} \label{eq:Nthree_result}
    R^3 (1+2X) & \textrm{for short delays,} \\
    R^3 (1+X) & \textrm{for long delays,}
  \end{cases} \\
  \hspace{-2ex}\textrm{and~}
  \Nsi(\tau) &=
  \begin{cases} \label{eq:Nsi_result}
    R^2 (1+X) & \textrm{for short delays,} \\
    R^2 & \textrm{for long delays,}
  \end{cases}
\end{align}
where ``long'' and ``short'' delays mean approximately $|\tau|>\tcoin + \tau_d$
and $|\tau|<\tcoin - \tau_d$ respectively. It is not surprising that only the
accidental-coincidence factor, $R^3$, and the signal-idler coincidence-peak
factor, $X=(2R\tcoin)^{-1}$, show up in the formula for $\Nthree$: all
signal-signal effects are in fact too small to be detectable unless explicitly
looked for. The graph of $\gcav(\tau)$, Eq.\ref{eq:defgcav}, is therefore
\begin{equation} \label{eq:gcav_result}
  \gcav(\tau) =
  \begin{cases}
    \sim \frac{1+2X}{(1+X)^2} \sim 4R\:\tcoin& 
  \textrm{for short delays} \\
  1 & \textrm{for long delays}.
  \end{cases}
\end{equation}
This conclusion is quantitatively confirmed by visual inspection of
\cite[Fig.~3 and Fig.~4]{Bocquillon09} (in the limit $\tau_c \gg \tau_d$ of
course\footnote{If $\tau_c\!\not\gg\!\tau_d$ but the convolution time scale is
  still much larger than $\Delta t$, then
  Eqs.\,\ref{eq:Nthree_result}-\ref{eq:gcav_result} hold valid with $X=p/R$,
  where $p$ is the height of the plateau of the convolution function.}).
Therefore, far from addressing the probability of detecting two signal photons
conditioned on the detection of an idler photons, these authors are again just
measuring the temporal correlation between the signal and idler arm.

\section*{Meaning of the conditioned second-order coherence}

Let us come back to the definition of second-order signal coherence {\em
  conditioned} on the detection of an idler photon at time $t_i$, given in
Eq.\ref{eq:def_gc}. This is the standard definition of Glauber's $\gtwo$, but
over a post-measurement state obtained by applying the idler measurement
operator to the unconditioned SPDC density matrix:
\begin{equation}
  \rho \longrightarrow \rho_{t_i} = \frac{E_i(t_i) \, \rho \, E_i^\dagger(t_i)}
  {\mathrm{tr}[\rho \, E_i^\dagger(t_i) E_i(t_i)]}.
\end{equation}
The resulting state is of course not time invariant, because of the
introduction of the parameter $t_i$. Equivalently, one can use modified
observables for the evaluation of expectation values on the original state:
\begin{equation}
  \langle Y \rangle_{\mathrm{pm}} \longrightarrow
  \frac{\langle E_i^\dagger(t_i) \,Y\, E_i(t_i) \rangle}
  {\langle E_i^\dagger(t_i) E_i(t_i) \rangle}.
\end{equation}

The usefulness of the $\gtwo(t_1,t_2)$ function for a {\em time-invariant}
field is that it is insensitive to detector efficiencies and depends only on
$\tau = t_2-t_1$, proportional to the probability (or rate, for a continuous
field) of finding two photons at times $t_1$ and $t_2$. If $\gtwo(\tau)$
decreases (increases) in some neighbourhood of $\tau=0$ the field is said to be
{\em bunched} {\em (anti-bunched)}, meaning that photons ``prefer'' to come
together (alone). If $\gtwo(0)$ is smaller (larger) than one the field
statistics in a sufficiently small time window is sub-Poissonian
(super-Poissonian).

However, due to the introduction of the idler time $t_i$, the conditioned SPDC
field is no more time invariant, and \smash{$\gc$} turns out not to be
proportional to the rate of arrival of signal photons at time $t_1$ and $t_2$,
because the normalisation factors in the denominator of Eq.\ref{eq:def_gc} are
not constant. For instance, if $t_2$ is set to a very large time, using
Eqs.\ref{eq:formula_gc}, \ref{eq:formula_Ratessi}, and \ref{eq:value_gsi} it is
immediate to find that
\begin{equation}
  \gc(t_1,t_2=\infty|t_i) = 1,
\end{equation}
irrespectively of the delay $t_1-t_i$, although the detection rate on the first
signal detector is much larger when $t_1=t_i$ with respect to the case $t_1-t_i
\gg \Delta t$, due to signal-idler correlation. Therefore $\gc$ is not a good
substitute for detection rates; one should instead directly use the probability
density of detecting signal photons at times $t_1$ and $t_2$ provided an idler
photon was detected at time $t_i$, which, due to Bayes' theorem, is just the
triple-coincidence rate $\Ratessi(t_1,t_2,t_i)$ divided by $R$. 

It is also to be remarked that the last two special cases described in
\cite[pag.3, right]{Bocquillon09}, which should presumably corroborate the
idea that $\gc$ is a sensible figure of merit, can also be obtained from
$\Ratessi$. Specifically, these examples concern the fact that $\Delta t$ is
the signal-idler ``coherence time'',
\begin{displaymath}
  \frac{\Ratessi(t_1,\infty | t_i) }
       {\Ratessi(-\infty,\infty | t_i) } 
       = 1 + \frac{C^2(t_1-t_i) }{ R^2} = \gsi(t_1-t_i),
\end{displaymath}
~\par\noindent and that unconditioned signal-signal events are thermal,
\begin{displaymath}
  \frac{\Ratessi(t_1,t_2 | -\infty) }
       {\Ratessi(t_1,\infty | -\infty) } 
       = 1 + \frac{R^2(t_1-t_2)}{R^2}
       = 1 + |g_s^{(1)}(t_1-t_2)|^2.
\end{displaymath}

On the other hand, instead of showing any sign of sub-Poissonian behaviour, as
the last special case $\gc(t_i,t_i|t_i) \sim 0$ seems to suggest, the point
$t_1=t_2=t_i$ is the maximum of $\Ratessi$. This is in agreement with the fact
that one-arm SPDC radiation is thermal: if an idler photon was detected at time
$t_i$ and a signal photon at time $t_1=t_i$, then the most likely time for a
second signal photon to appear is just $t_2=t_i$.

\section*{Conclusions}

It is shown that the experimental findings reported in \cite[Fig.~3
  and~4]{Bocquillon09} about conditioned signal-signal coincidences in an SPDC
source are indeed related not to the conditioned second-order signal-signal
coherence, but to the much simpler (to measure) and stronger signal-idler
correlation. The author acknowledges fruitful discussions on the subject with
H. H\"ubel and I. M. Herbauts.


\begin{thebibliography}{99}
\bibitem{Blauensteiner09} B.~Blauensteiner, I.~Herbauts, S.~Bettelli, A.~Poppe,
  and H.~H\"ubel, {\em Photon bunching in parametric down-conversion with
    continuous wave excitation}, Phys. Rev. A {\bf 79}, 063846. % (2009).
\bibitem{Bocquillon09} E.~Bocquillon, C.~Couteau, M.~Razavi, R.~Laflamme, and
  G.~Weihs, {\em Coherence measures for heralded single-photon sources},
  Phys. Rev. A {\bf 79}, 035801 (2009). %, arXiv:0807.1725v2.
\bibitem{Razavi09} M.~Razavi, I.~S\"ollner, E.~Bocquillon, C.~Couteau,
  R.~La\-flamme, and G.~Weihs, {\em Characterizing heralded single-photon
    sources with imperfect measurement devices}, J. Phys. B:
  At. Mol. Opt. Phys. {\bf 42}, 114013 (2009). %,arXiv:0812.2445.
\end{thebibliography}
\end{document}